\documentclass[twocolumn,aps,prd,10pt,superscriptaddress,longbibliography, floatfix,citeautoscript]{revtex4-2}

\usepackage{amsmath}
\usepackage{bm}
\usepackage{glossaries}
\usepackage{graphicx}
\usepackage{multirow}
\usepackage{threeparttable}
\usepackage[usenames,dvipsnames,svgnames]{xcolor}
\usepackage{hyperref}
\hypersetup{
pdfnewwindow=true,      
colorlinks=true,        
linkcolor=Blue,         
citecolor=Blue,         
filecolor=Blue,         
urlcolor=Blue           
}

\newacronym{blackP}{Black-P}{black phosphorene}
\newacronym{blueP}{Blue-P}{blue phosphorene}
\newacronym{violetP}{Violet-P}{violet phosphorene}
\newacronym{2D}{2D}{two-dimensional}
\newacronym{md}{MD}{molecular dynamics}
\newacronym{mlp}{MLP}{machine-learned potential}
\newacronym{gap}{GAP}{Gaussian approximation potential}
\newacronym{nep}{NEP}{neuroevolution potential}
\newacronym{nn}{NN}{neural network}
\newacronym{snes}{SNES}{separable natural evolution strategy}
\newacronym{dft}{DFT}{density functional theory}
\newacronym{mbd}{MBD}{many-body dispersion}
\newacronym{rmse}{RMSE}{root mean square error}
\newacronym{hnemd}{HNEMD}{homogeneous non-equilibrium molecular dynamics}
\newacronym{mfp}{MFP}{mean free path}
\newacronym{bte}{BTE}{Boltzmann transport equation}
\newacronym{ald}{ALD}{anharmonic lattice dynamics}

\begin{document}

\title{Variable thermal transport in black, blue, and violet phosphorene from extensive atomistic simulations with a neuroevolution potential}

\author{Penghua Ying}
\affiliation{School of Science, Harbin Institute of Technology, Shenzhen, 518055, P. R. China}

\author{Ting Liang}
\affiliation{Department of Electronic Engineering and Materials Science and Technology Research Center, The Chinese University of Hong Kong, Shatin, N.T., Hong Kong SAR, 999077, P. R. China}

\author{Ke Xu}
\affiliation{Department of Physics, Research Institute for Biomimetics and Soft Matter, Jiujiang Research Institute and Fujian Provincial Key Laboratory for Soft Functional Materials Research, Xiamen University, Xiamen 361005, PR China.}

\author{Jin Zhang}
\affiliation{School of Science, Harbin Institute of Technology, Shenzhen, 518055, P. R. China}

\author{Jianbin Xu}
\affiliation{Department of Electronic Engineering and Materials Science and Technology Research Center, The Chinese University of Hong Kong, Shatin, N.T., Hong Kong SAR, 999077, P. R. China}

\author{Jianyang Wu}
\affiliation{Department of Physics, Research Institute for Biomimetics and Soft Matter, Jiujiang Research Institute and Fujian Provincial Key Laboratory for Soft Functional Materials Research, Xiamen University, Xiamen 361005, PR China.}

\author{Zheyong Fan}
\email{brucenju@gmail.com}
\affiliation{College of Physical Science and Technology, Bohai University, Jinzhou 121013, P. R. China}
\affiliation{MSP group, QTF Centre of Excellence, Department of Applied Physics, Aalto University, FI-00076 Aalto, Espoo, Finland}

\author{Tapio Ala-Nissila}
\affiliation{MSP group, QTF Centre of Excellence, Department of Applied Physics, Aalto University, FI-00076 Aalto, Espoo, Finland}
\affiliation{Interdisciplinary Centre for Mathematical Modelling, Department of Mathematical Sciences, Loughborough University, Loughborough, Leicestershire LE11 3TU, UK}

\author{Zheng Zhong}
\email{zhongzheng@hit.edu.cn}
\affiliation{School of Science, Harbin Institute of Technology, Shenzhen, 518055, P. R. China}
\date{\today}

\begin{abstract}
Phosphorus has diverse chemical bonds and even in its two-dimensional form there are three stable allotropes: black phosphorene (Black-P), blue phosphorene (Blue-P), and violet phosphorene (Violet-P). Due to the complexity of these structures, no efficient and accurate classical interatomic potential has been developed for them. In this paper, we develop an efficient machine-learned neuroevolution potential model for these allotropes and apply it to study thermal transport in them via extensive molecular dynamics (MD) simulations. Based on the homogeneous nonequilibrium MD method, the thermal conductivities are predicted to be $12.5 \pm 0.2$ (Black-P in armchair direction), $78.4 \pm 0.4$ (Black-P in zigzag direction), $128 \pm 3$ (Blue-P), and $2.36 \pm 0.05$ (Violet-P) $\mathrm{Wm^{-1}K^{-1}}$. The underlying reasons for the significantly different thermal conductivity values in these allotropes are unraveled through spectral decomposition, phonon eigenmodes, and phonon participation ratio. Under external tensile strain, the thermal conductivity in black-P and violet-P are finite, while that in blue-P appears unbounded due to the linearization of the flexural phonon dispersion that increases the phonon mean free paths in the zero-frequency limit.
\end{abstract}
\maketitle

\section{Introduction}

\begin{figure}[htb]
\begin{center}
\includegraphics[width=\columnwidth]{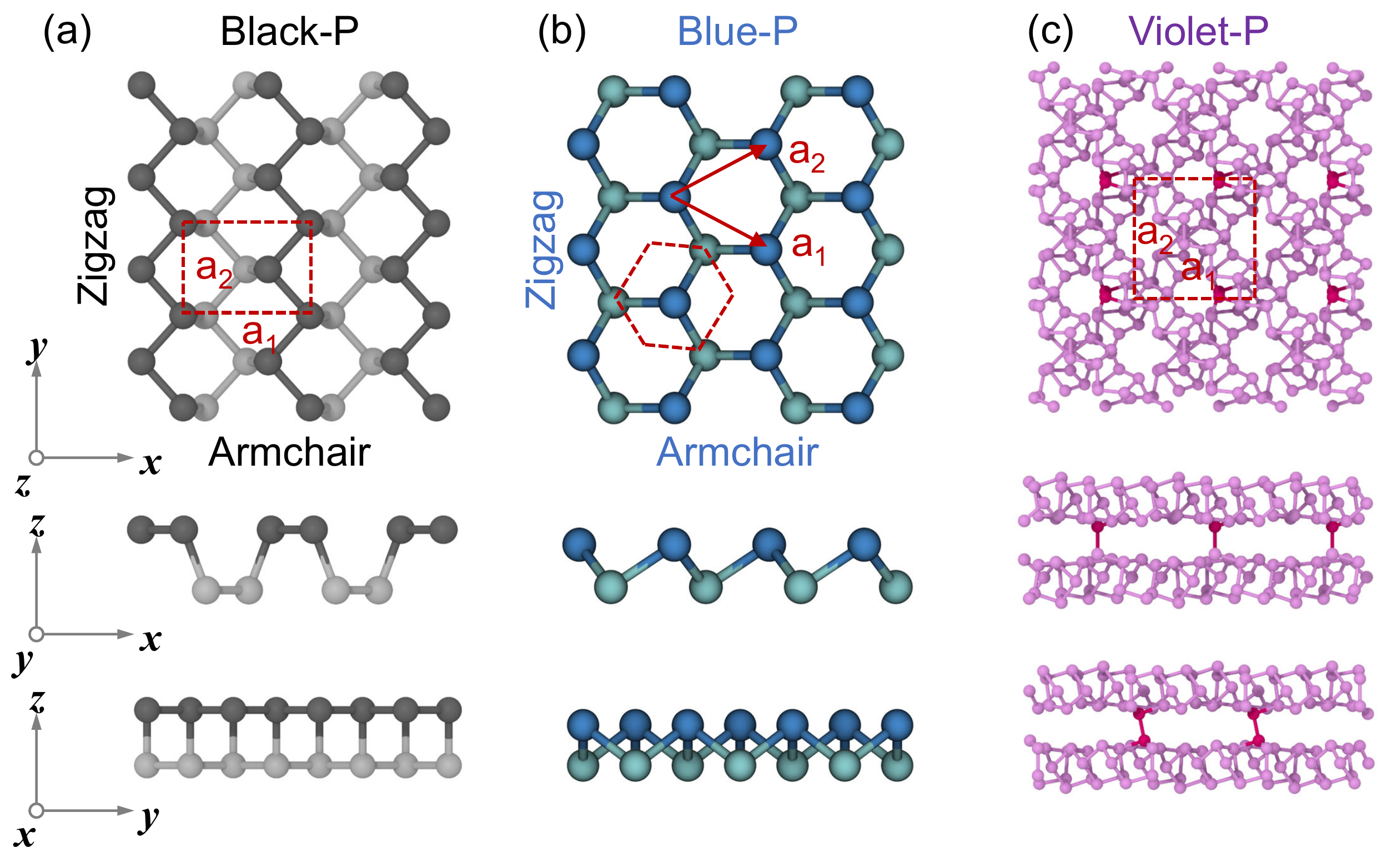}
\caption{Crystal structures of (a) \gls{blackP} (b) \gls{blueP} and (c) \gls{violetP}. The in-plane primitive cell vectors $a_1$ and $a_2$ are indicated. \gls{blackP} and \gls{blueP} have two distinct directions that are referred to as the armchair and zigzag directions chosen to be aligned with the $x$ and $y$ axes, respectively. The \textsc{ovito}  package \cite{stukowski2009visualization} is used for visualization.}
\label{figure:model}
\end{center}
\end{figure}

Phonon thermal transport in \gls{2D} materials exhibits unique properties that are absent in the bulk form \cite{gu2018rmp}. After the discovery of the high thermal conductivity of graphene \cite{balandin2008superior}, thermal transport in other \gls{2D} materials has been actively studied. Among them, \gls{2D} phosphorene occupies a special role as there are three stable  allotropes: \gls{blackP}, \gls{blueP}, and \gls{violetP} \cite{carvalho2016phosphorene,zhang2020structure}. These allotropes have very different crystalline structures: \gls{blackP} has an anisotropic orthorhombic structure with four atoms in the primitive cell (\autoref{figure:model}(a)); \gls{blueP} has a honeycomb structure buckled perpendicular to the 2D plane with two atoms in the primitive cell (\autoref{figure:model}(b)); while \gls{violetP} has a very complicated tubular structure with 42 atoms in the primitive cell (\autoref{figure:model}(c)). All the three allotropes are semiconductors with direct electronic bandgaps \cite{tran2014layer,Zhu2014PRL,zhang2020structure}. Therefore, phonons are the major heat carriers and phonon-dominated thermal transport in them is of great interest.

Precise measurement of the thermal conductivity $\kappa$ of \gls{2D} materials is quite challenging. For \gls{blackP}, only multi-layer samples thicker than 5 nm have been measured \cite{Jang2015AM, Luo2015NC} and there are so far no measurements for the other two allotropes. In this regard, computational methods play an important role in characterizing the phonon-mediated thermal transport properties in these allotropes. \gls{bte}, combined with \gls{ald} has been the major tool for studying heat transport in various materials \cite{lindsay2016first, mcgaughey2019phonon, bao2018review}. There have been quite a few \gls{bte}-\gls{ald} studies for both \gls{blackP} \cite{Jain2015SR,Zhu2014PRB,Qin2015PCCP,Liu2015Nanoscale,Zheng2016PRB,Zhang2017SR} and \gls{blueP} \cite{Jain2015SR,Zheng2016PRB,Zhang2017SR,Liu2018APL,Zhu2020PhysicaE}. However, the \gls{bte}-\gls{ald} method quickly becomes computationally infeasible when the primitive cell contains a large number of atoms, which explains the absence of the application of it to \gls{violetP}. Perhaps the most feasible method for complex crystals is based on \gls{md} simulations which also account for phonon anharmonicity to arbitrary order and contain phonon coherence effects \cite{zhang2022heat}. However, the application of \gls{md} simulation to phonon thermal transport in the  phosphorene allotropes has been hindered by the lack of accurate, efficient and transferable interatomic potentials. 

Recently, \glspl{mlp} have been demonstrated to be a promising approach to enable reliable \gls{md} simulations with a reasonable computational cost \cite{behler2016perspective, deringer2019machine}. A \gls{gap} \cite{bartok2010gaussian} for many phosphorus allotropes has already been developed \cite{Deringer2020NC} and has been shown to be very accurate and transferable, but the efficiency of this \gls{gap} model is currently not high enough to be used to study phonon thermal transport in the phosphorene allotropes \textit{via} \gls{md} simulations. Another \gls{mlp} framework developed by some of the present authors called the \gls{nep} \cite{Fan2021PRB, fan2022improving, fan2022GPUMD} has been shown to have much higher computational efficiency while achieving comparable accuracy. In this paper, we use part of the training data as used for the \gls{gap} model \cite{Deringer2020NC} to train an accurate and highly efficient \gls{nep} model and use it to comprehensively characterize the versatile and highly varying phonon thermal transport properties of the 2D phosphorene allotropes, revealing the underlying phonon transport mechanisms and the crucial roles played by external strain in regularizing the heat transport in these materials. 

\section{Training and validating a NEP model for phosphorene \label{section:NEP}}

\gls{nep} \cite{Fan2021PRB,fan2022improving,fan2022GPUMD} is a type of \gls{mlp} based on a \gls{nn} and is trained using the \gls{snes} \cite{Schaul2011}. A number of atom-environment descriptor components of a central atom are used as the input layer of the \gls{nn} and the energy of the central atom is taken as the output of the \gls{nn}, which is the same as in the standard Behler-Parrinello high-dimensional \gls{nn} potential  \cite{behler2007prl}. The site energy $U_{i}$ of atom $i$ is taken as a function of $N_\mathrm{des}$ descriptor components. To model this function, a feedforward \gls{nn} with a single hidden layer with $N_\mathrm{neu}$ neurons is applied:
\begin{equation}
\label{equation:Ui}
U_i = \sum_{\mu=1}^{N_\mathrm{neu}}w^{(1)}_{\mu}\tanh\left(\sum_{\nu=1}^{N_\mathrm{des}} w^{(0)}_{\mu\nu} q^i_{\nu} - b^{(0)}_{\mu}\right) - b^{(1)},
\end{equation}
where $\mathbf{w}^{(0)}$, $\mathbf{w}^{(1)}$, $\mathbf{b}^{(0)}$, and $b^{(1)}$ are the trainable weights and bias parameters in the \gls{nn} and $\tanh(x)$ is the activation function. 

The descriptor consists of both radial and angular components and is constructed based on Chebyshev and Legendre polynomials (or spherical harmonics via the addition theorem), inspired by previous works \cite{behler2007prl,Caro2019prb}. For a central atom $i$, there is a set of radial descriptor components $\{q^i_{n}\}$ ($0\leq n\leq n_\mathrm{max}^\mathrm{R}$): 
\begin{equation}
\label{equation:qin}
q^i_{n}
= \sum_{j\neq i} g_{n}(r_{ij}),
\end{equation}
and a set of angular descriptor components  $\{q^i_{nl}\}$ ($0\leq n\leq n_\mathrm{max}^\mathrm{A}$ and $1\leq l \leq l_\mathrm{max}$):
\begin{equation}
q^i_{nl} 
= \frac{2l+1}{4\pi}\sum_{j\neq i}\sum_{k\neq i} g_n(r_{ij}) g_n(r_{ik})
P_l(\cos\theta_{ijk}).
\label{equation:qinl_legendre}
\end{equation}
Here the summation runs over all the neighbors of atom $i$ within a certain cutoff distance. $P_l(\cos\theta_{ijk})$ is the Legendre polynomial of order $l$ and $\theta_{ijk}$ is the angle formed by the $ij$ and $ik$ bonds. 

The radial function $g_n(r_{ij})$ in \autoref{equation:qin} and \autoref{equation:qinl_legendre} are defined as:
\begin{equation}
g_n(r_{ij}) = \frac{c_{nij}}{2}
\left[
    T_n\left(2\left(r_{ij}/r_\mathrm{c}-1\right)^2-1\right)+1
\right]
f_\mathrm{c}(r_{ij}).
\label{equation:f_n}
\end{equation}
Here, $T_n(x)$ is the $n^{\rm th}$ order Chebyshev polynomial of the first kind and $f_\mathrm{c}(r_{ij})$ is the cutoff function defined as
\begin{equation}
   f_\mathrm{c}(r_{ij}) 
   = \begin{cases}
   \frac{1}{2}\left[
   1 + \cos\left( \pi \frac{r_{ij}}{r_\mathrm{c}} \right) 
   \right],& r_{ij}\leq r_\mathrm{c}; \\
   0, & r_{ij} > r_\mathrm{c}.
   \end{cases}
\end{equation}
The cutoff radius $r_{\rm c}$ for the radial and angular descriptor components can be different, and are denoted as $r_{\rm c}^{\rm R}$ and $r_{\rm c}^{\rm A}$, respectively. The coefficients $c_{nij}$ are trainable parameters that depend on the types of the atoms $i$ and $j$. The choice of the hyperparameters $r_{\rm c}^{\rm R}$, $r_{\rm c}^{\rm A}$, $n_{\rm max}^{\rm R}$, $n_{\rm max}^{\rm A}$, and $l_{\rm max}$ will be discussed below. The total number of descriptor components is $N_{\rm des} = (n_{\rm max}^{\rm R}+1) + (n_{\rm max}^{\rm A}+1) l_{\rm max}$.

The training data we used consist of the \gls{2D} phosphorene structures as constructed by Deringer \textit{et al.} \cite{Deringer2020NC,gap_train} and $60$ extra ones ($20$ for each of the three allotropes) with $0.1$ \AA~ random  displacements for each atom starting from the equilibrium structures. For the $60$ extra structures, we performed \gls{dft} calculations that are consistent with those in Ref. \cite{Deringer2020NC}, using the Perdew-Burke-Ernzerhof functional \cite{Perdew1996PRL} plus the \gls{mbd} \cite{Tkatchenko2012PRL,Ambrosetti2014JCP}, and the projector-augmented wave method \cite{Blchl1994PRB} as implemented in \textsc{vasp} \cite{Kresse1996PRB,Kresse1999PRB}. All the calculations were converged with an energy tolerance of 10$^{-8}$ eV under an energy cutoff of 500 eV. Finally, our training data set contains $2139$ structures ($51191$ atoms in total) including nanoribbons, \gls{2D} structures, and bulk structures, and each structure has energy, force, and virial data. The trained \gls{nep} model is tested against a hold-out dataset consisting of $309$ \gls{2D} phosphorene structures ($3468$ atoms in total) \cite{gap_test}.     

The \gls{nep} model was trained using the \textsc{gpumd} package (version $3.3.1$) \cite{fan2017cpc, fan2022GPUMD}, choosing the NEP2 version. The cutoff radii for the radial and angular descriptor components are $r_{\rm c}^{\rm R}=8$ \AA~ and $r_{\rm c}^{\rm A}=5$ \AA, respectively. We note that with a large radial cutoff distance of $r_{\rm c}^{\rm R}=8$ \AA, we do not need to include an empirical dispersion interaction term (such as a Lennard-Jones potential) explicitly to account for the van der Waals interactions. The Chebyshev polynomial expansion order for the radial and angular descriptor components are $n_{\rm max}^{\rm R}=15$ and $n_{\rm max}^{\rm A}=10$, respectively. The Legendre polynomial expansion order for the angular descriptor components is $l_{\rm max}=4$. The number of neurons in the hidden layer of the \gls{nn} is $N_\mathrm{neu}=40$. 

A loss function to be minimized is defined as follows:
\begin{equation}
L = \lambda_1 L_1 + \lambda_2 L_2 + \lambda_{\rm e} \Delta U + \lambda_{\rm f} \Delta F + \lambda_{\rm v} \Delta W
\end{equation}
where $\Delta U$, $\Delta F$, and $\Delta W$ are the \glspl{rmse} of energy, force, and virial, respectively, between the predicted and the reference values, $L_1$ and $L_2$ are proportional to the 1-norm and 2-norm of the training parameters, and $\lambda_{\rm e}$, $\lambda_{\rm f}$, $\lambda_{\rm v}$, $\lambda_{1}$, $\lambda_{2}$ are the weights of the various terms. We choose $\lambda_{1}=\lambda_{2}=0.05$, $\lambda_{\rm e}=\lambda_{\rm f}=1$, and $\lambda_{\rm v}=0.1$. The population size and number of generations in the \gls{snes} algorithm \cite{Schaul2011} are $N_{\rm pop}=50$ and $N_{\rm gen}=2 \times 10^5$. 

\begin{figure}[htb]
\begin{center}
\includegraphics[width=\columnwidth]{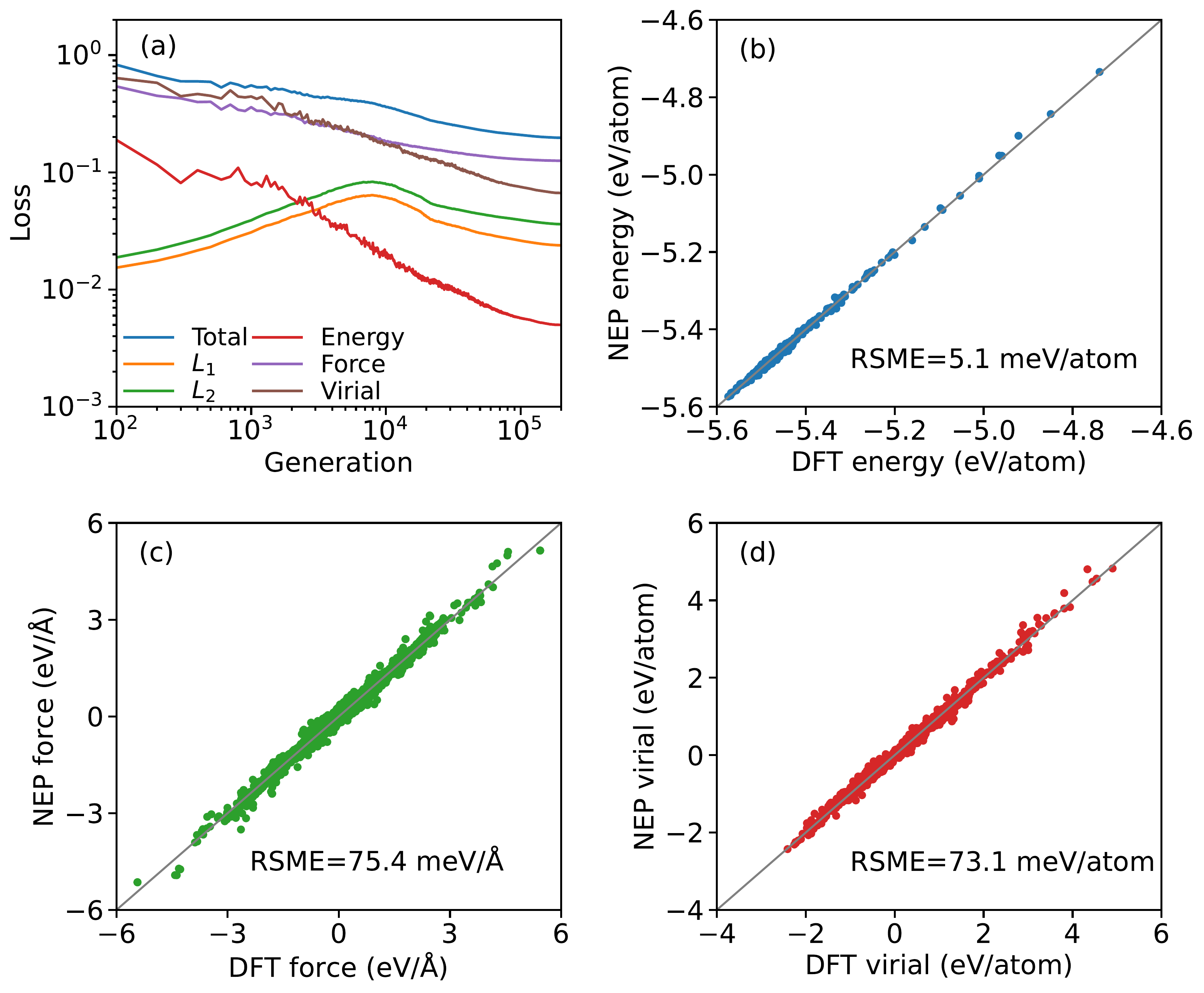}
\caption{(a) Evolution of various terms in the loss function for the training data set with respect to the generation in the \gls{snes} algorithm. (b) Energy, (c) force, and (d) virial calculated from \gls{nep} as compared to the reference data for the testing data set. The lines in (b)-(d) represent the identity function used to guide the eyes.}
\label{figure:train}
\end{center}
\end{figure}

\begin{figure}[htb]
\begin{center}
\includegraphics[width=\columnwidth]{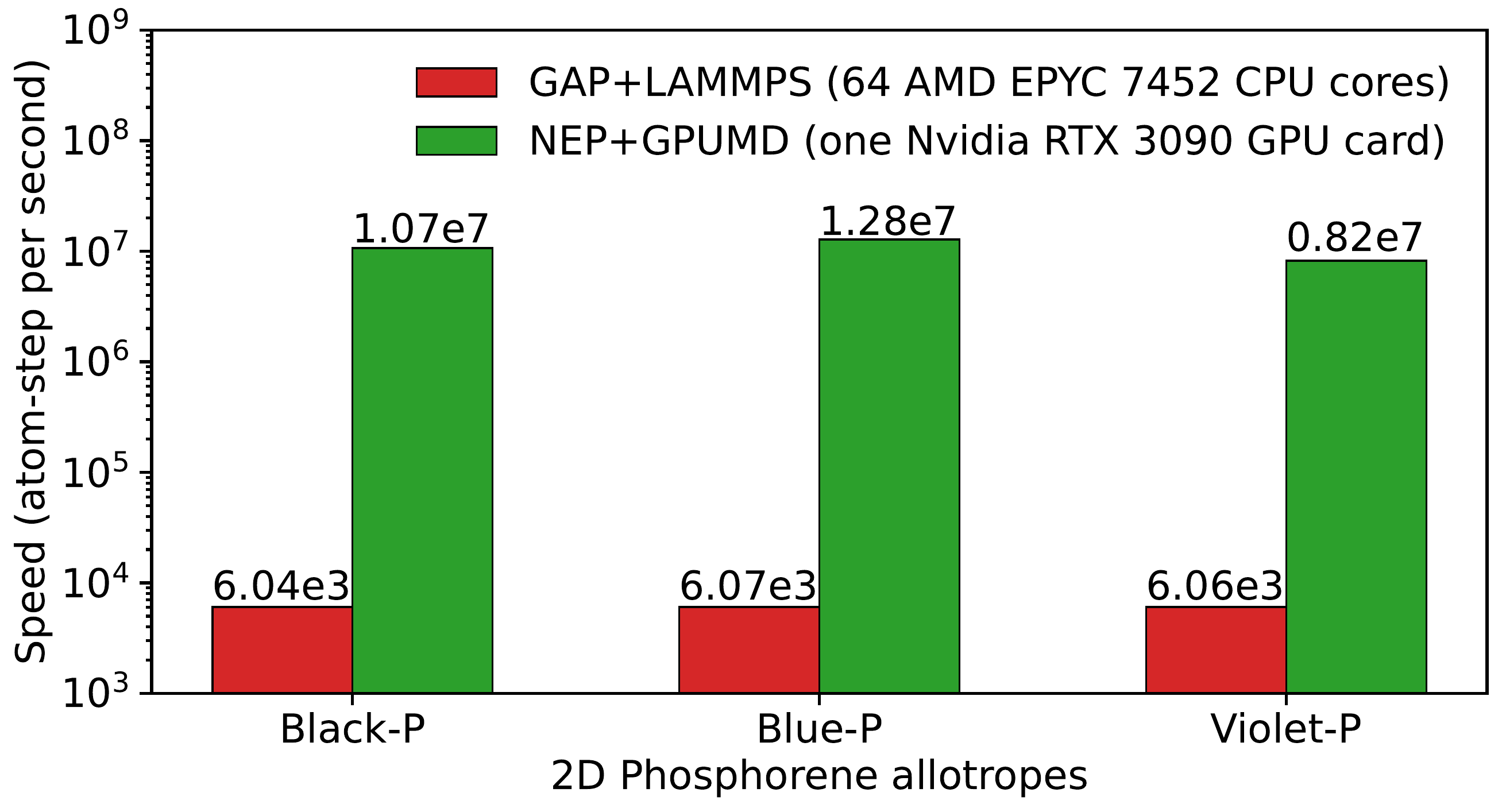}
\caption{A comparison of the computational speed of \gls{nep}-GPUMD (running with an Nvidia RTX 3090 GPU card) and \gls{gap}-LAMMPS (running with 64 AMD EPYC 7452 CPU cores) for the \gls{2D} phosphorene allotropes. We note that there are two versions of \gls{gap} in Ref. \cite{Deringer2020NC}, and we used the faster one without the dispersion part to give a fair comparison with our \gls{nep} model.}
\label{figure:speed}
\end{center}
\end{figure}

In \autoref{figure:train}(a) we show the evolution of various terms in the loss function with respect to the \gls{snes} generation during the training process. With $N_{\rm gen}=2\times10^{5}$ generations, the \glspl{rmse} of energy, force, and virial are essentially converged. In \autoref{figure:train} (b)-(d) we compare the energy, force, and virial predicted by the \gls{nep} model and those from \gls{dft} calculations for the testing data set. The RSMEs of energy, force, and virial are $5.1$ meV/atom, $75.4$ meV/\AA, and $73.1$ meV/atom, respectively. As a comparison, we note that the \glspl{rmse} of energy and force from \gls{gap} (virial \gls{rmse} is not available) for the \gls{2D} structures are $2.0$ meV/atom and  $70.0$ meV/\AA, respectively \cite{Deringer2020NC}. 

\gls{nep} as implemented in \textsc{gpumd} attains a much higher computational performance than \gls{gap} \cite{Deringer2020NC} as implemented in \textsc{quip} \cite{quip} and interfaced with LAMMPS (version 14Dec2021) \cite{thompson2022lammps}. The computational speed is measured by running \gls{md} simulations for 100 steps in the isothermal ensemble, using $19200$, $14040$, and $32928$ atoms for \gls{blackP}, \gls{blueP}, and \gls{violetP}, respectively. From \autoref{figure:speed}, we see that the computational speed of \gls{nep} using a single Nvidia RTX $3090$ GPU card is of the order of $10^{7}$ atom-step per second, which is more than three orders of magnitude higher than that of \gls{gap} using $64$ AMD EPYC $7452$ CPU cores (two nodes, each with $32$ cores). The high computational efficiency of \gls{nep} is crucial for its application to thermal transport calculations, which require extensive sampling of the phase space. To be exact, the whole \gls{md} simulations in this paper took about 1000 hours using a single Nvidia RTX $3090$ GPU card.

\begin{table}[htb]
\centering
\setlength{\tabcolsep}{1Mm}
\caption{A comparison between lattice constants of \gls{2D} phosphorene allotropes predicted by our \gls{nep} model and \gls{dft}-\gls{mbd} calculations.}
\label{table:lattice_compare}
\begin{tabular}{lllllll}
\hline
\hline
Allotrope      & \multicolumn{2}{l}{Black} & \multicolumn{2}{l}{Blue} & \multicolumn{2}{l}{Violet} \\
\hline
Lattice constant (\AA) & $a_{1}$ & $a_{2}$ & $a_{1}$ & $a_{2}$ & $a_{1}$ & $a_{2}$ \\
\hline
\gls{nep}               &  4.39  & 3.30   & 3.26   & 3.26   & 9.11   & 9.18   \\
\hline
\gls{dft}-\gls{mbd}         &  4.33  & 3.31   & 3.26   & 3.26   & 9.12   & 9.18   \\
\hline
\hline
\end{tabular}
\end{table}

\autoref{table:lattice_compare} compares the lattice constants of \gls{blackP}, \gls{blueP}, and \gls{violetP} (see \autoref{figure:model} for the definitions of the lattice constants) predicted by \gls{nep} with those predicted by \gls{dft}-\gls{mbd} calculations. Our \gls{nep} model can predict the lattice constants of the \gls{2D} phosphorene allotropes very well, with a relative error being of the order of $0.1\%$ in most cases and $1\%$ for $a_1$ in \gls{blackP}.

\begin{figure*}[htb]
\begin{center}
\includegraphics[width=1.4\columnwidth]{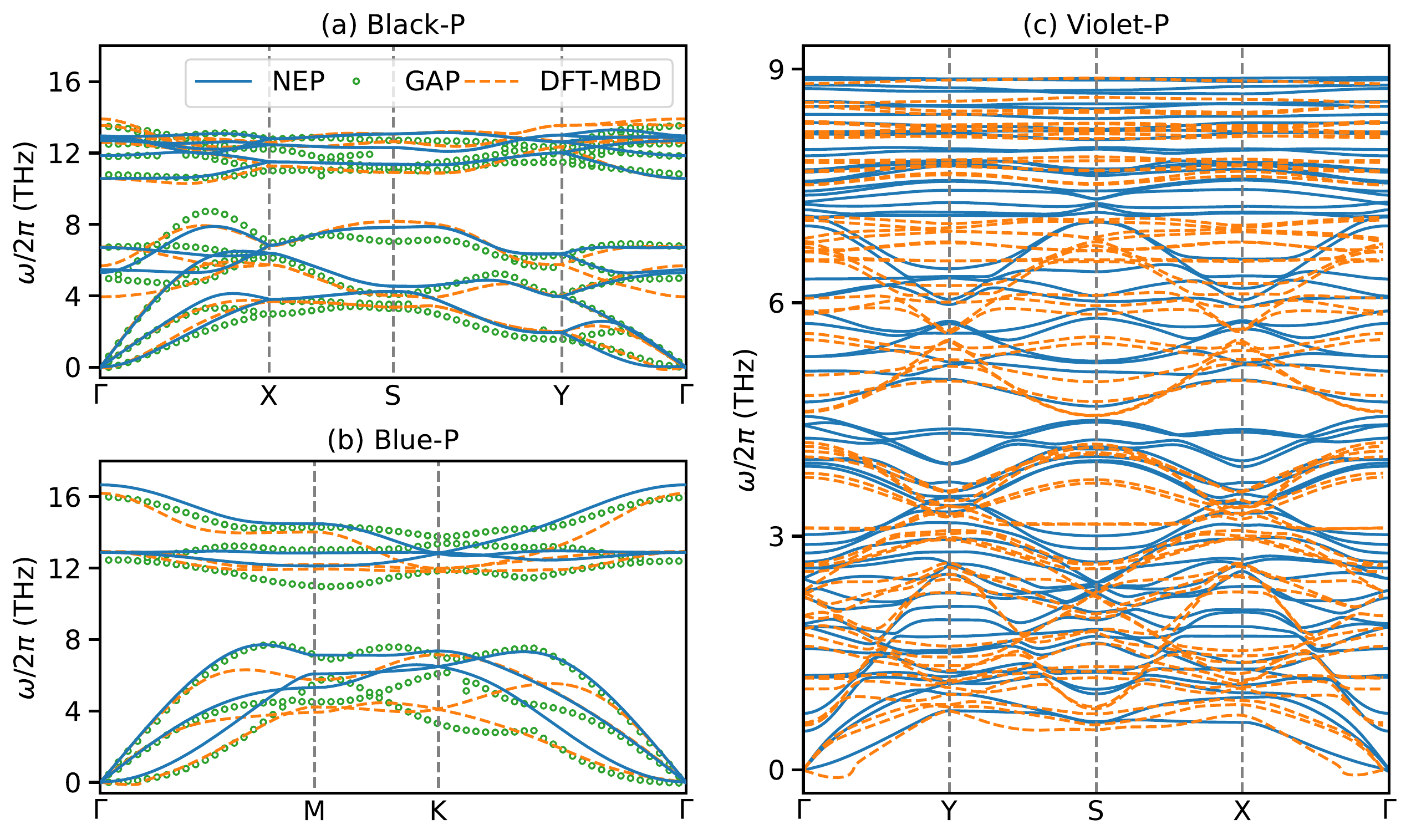}
\caption{Phonon dispersions of (a) \gls{blackP}, (b) \gls{blueP}, and (c) \gls{violetP} from the \gls{nep} model (solid lines), \gls{gap} model (circles), and \gls{dft}-\gls{mbd} calculations (dashed lines). The \gls{gap} results are from Ref.~\cite{koneru2022multi}. Our \gls{dft}-\gls{mbd} results for \gls{blackP} and \gls{blueP} closely match those reported by Zhu \textit{et al.} \cite{Zhu2014PRL} and Jain \textit{et al.} \cite{Jain2015SR}. The \textsc{phonopy} package \cite{togo2015first} is used for obtaining the \gls{dft}-\gls{mbd} results.} 
\label{figure:phonon}
\end{center}
\end{figure*}

The primitive cells of \gls{blackP}, \gls{blueP}, and \gls{violetP} are 4, 2, and 42, respectively, leading to 12, 6, and 126 phonon branches, respectively. Despite the very different phonon dispersions, our \gls{nep} model can well describe them simultaneously, exhibiting a level of accuracy similar to the GAP model, as can be seen from \autoref{figure:phonon}. This is beyond the reach of any current empirical potential. For \gls{blueP}, the frequencies around the M and K points are overestimated as compared to \gls{dft}-\gls{mbd}, but those around the $\Gamma$ point are well described. The complex phonon dispersions of \gls{violetP} are also reasonably described by \gls{nep}, which even behaves more nicely around the $\Gamma$ point than \gls{dft}-\gls{mbd} which suffers from some numerical issues. 

We note that there is a phonon band gap with $2.6$ THz ($7.9$ to $10.5$ THz) in \gls{blackP} and one with $4.6$ THz ($7.7$ to $12.3$ THz) in \gls{blueP}, while there is no evident gap in \gls{violetP}. In the case of \gls{violetP} which has a large unit cell, there are flat bands at lower frequencies, and significant overlap between multiple bands in which case approaches based on the linearized BTE are expected to fail \cite{simoncelli2019unified}. Based on the band picture we expect a high degree of phonon localization in \gls{violetP} that should lead to a low value of 
thermal conductivity.

Before detailed calculations, we can also infer some interesting properties based on the acoustic branches. In \gls{blackP}, the acoustic branches are much higher in the $\Gamma$-X (zigzag) direction than in the $\Gamma$-Y (armchair) direction, leading to higher group velocities in the zigzag direction, which is one of the origins of the highly anisotropic phonon transport in \gls{blackP}. In the other two allotropes, there is no such anisotropy. As mentioned above, the acoustic branches of \gls{violetP} are much flatter than the other two allotropes, leading to  much lower phonon group velocities. This is also related to the flexibility of \gls{violetP} under deformation, which is caused by a unique deformation mode, namely, rotation of sub-nano rods \cite{ying2022tension}.

\section{Phonon thermal transport in  phosphorene allotropes \label{section:thermal}}

For a quantitative study we computed the thermal conductivity using the \gls{hnemd} method \cite{Evans1982PLA,Fan2019PRB} which has been shown to be an efficient approach in various \gls{2D} materials \cite{dong2018heat, xu2018thermal, gabourie2020reduced, wu2021thermal, kim2021extremely, wang2021anomalous, Ying2022IJHMT, sha2022thermal}. In this method, an external force is applied to each atom $i$ to drive the system out of equilibrium. The external force $\bm{F}_{i}^{\rm ext}$ can be written in terms of the per-atom energy $E_{i}$ and virial tensor $\textbf{W}_{i}$ as follows \cite{Gabourie2021PRB}: 
\begin{equation}
\label{equation:Fe}
\bm{F}_{i}^{\rm ext} = E_{i}\bm{F}_{\rm e} + \bm{F}_{\rm e}\cdot\textbf{W}_{i},
\end{equation}
where $\bm{F}_{\rm e}$ is the driving force parameter with the dimension of inverse length. The driving force will induce a nonequilibrium heat current $J(t)$ as a function of time $t$ that is linearly proportional to $F_{\rm e} = |\bm{F}_{\rm e}|$,
\begin{equation}
\label{equation:kappa}
\kappa (t) = \frac{J(t)}{TVF_{\rm e}},
\end{equation}
where $\kappa (t)$ is the instant thermal conductivity, $V$ is the volume, and $T$ is the temperature. In the calculation of $V$, the thicknesses of \gls{blackP} \cite{akai1989crystal}, \gls{blueP} \cite{Zhu2014PRL}, and \gls{violetP} \cite{zhang2020structure} are respectively taken  as 5.25 \AA, 5.63 \AA, and 11.00 \AA. The temperature needs to be maintained by using a thermostat, and to this end, we  use the Nos\'{e}-Hoover chain thermostat with a relaxation time of 100 fs. In all our \gls{md} simulations, a time step of 1 fs is used. The in-plane simulation cell size is set to 25 nm $\times$ 25 nm which is sufficiently large based on our tests.

To check the convergence of $\kappa (t)$, it is conventional to redefine the thermal conductivity as \cite{Fan2019PRB}:
\begin{equation}
\label{equation:kappa_cumulative}
\kappa (t) = \frac{1}{t}\int_{0}^{t}\frac{ J(\tau)}{TVF_{\rm e}}d\tau.
\end{equation}
The magnitude of the driving force parameter $F_{\rm e}$ should be small enough to keep the system within the linear response regime and large enough to keep a reasonably large signal-to-noise ratio. An upper bound of $F_{\rm e} = 1/\lambda_{\rm{max}}$ has been suggested \cite{Fan2019PRB}, where $\lambda_{\rm{max}}$ is the maximum phonon mean free path in the system, as will be further confirmed below.   

\begin{figure}[htb]
\begin{center}
\includegraphics[width=\columnwidth]{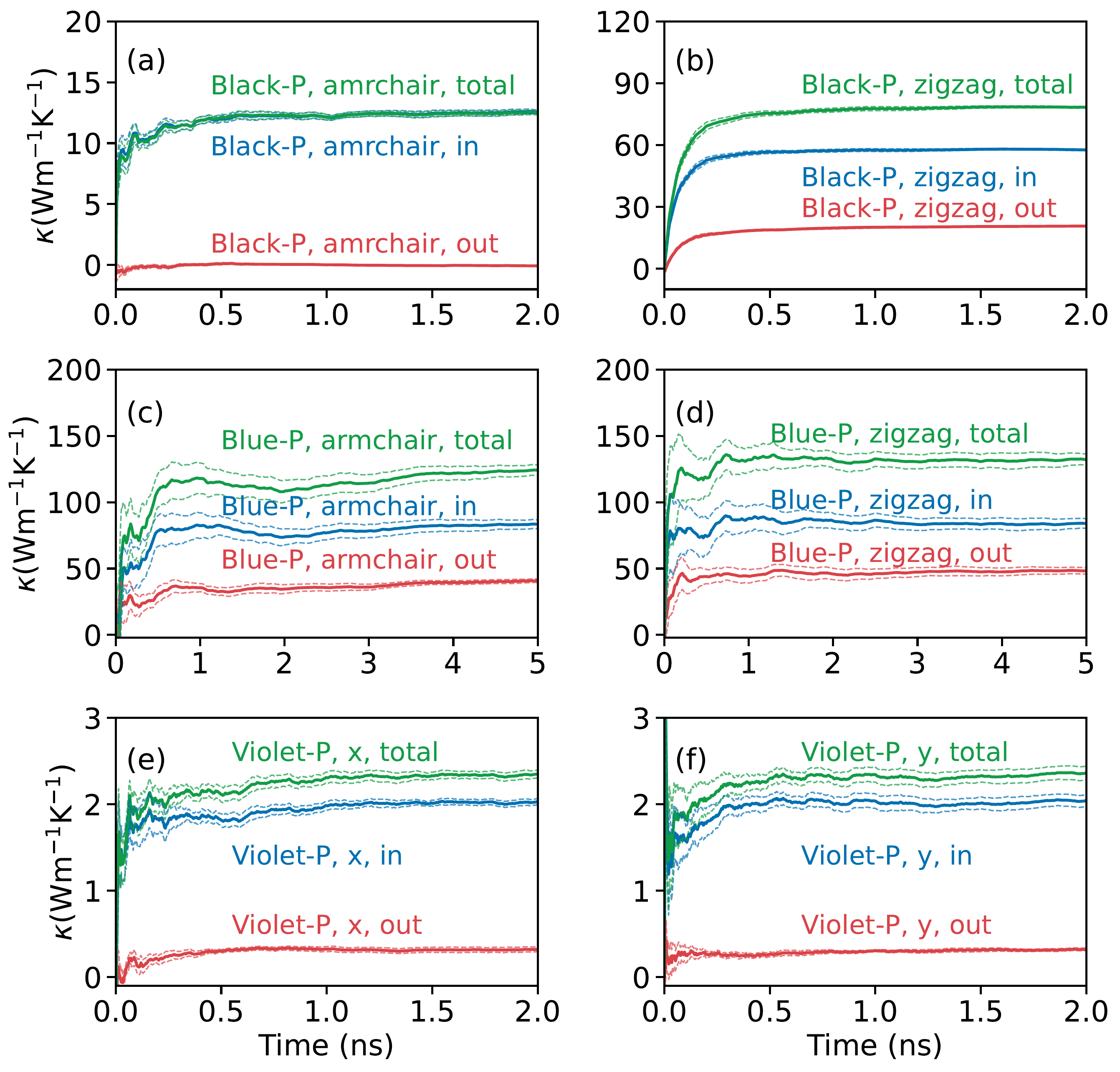}
\caption{(a)-(f) The thermal conductivity as defined in \autoref{equation:kappa_cumulative} for the phosphorene allotropes at $300$ K and zero pressure, along the zigzag/armchair or $x$/$y$ directions as defined in \autoref{figure:model}. In each subplot, the total thermal conductivity (``total'') is decomposed into contributions from in-plane (``in'') and out-of-plane (``out'') phonon modes. Solid lines are averaged values and dashed ones represent the error bounds from ten independent runs.} 
\label{figure:hnemd}
\end{center}
\end{figure}

For the $\kappa(t)$ obtained by using \autoref{equation:kappa_cumulative} in \gls{hnemd} simulations, we set $F_{\rm e}=0.1\ \rm{\mu m}^{-1}$ for \gls{blueP} (\autoref{figure:hnemd}(c)-(d)) and $F_{\rm e}=1.0\ \rm{\mu m}^{-1}$ for \gls{blackP} (\autoref{figure:hnemd}(a)-(b)) and \gls{violetP} (\autoref{figure:hnemd}(e) -(f)). We performed ten independent runs, each with a 100 ps equilibration stage in the isothermal-isobaric ensemble (with a target in-plane pressure of zero and a target temperature of 300 K), and after that, a 2-ns production stage for black and \gls{violetP}, or a 5-ns production stage for \gls{blueP}.  

In \autoref{figure:hnemd}, we decompose \cite{Fan2019PRB} $\kappa$ in to contributions form in-plane ($\kappa_{\rm in}$) and out-of-plane (flexural) ($\kappa_{\rm out}$) phonon modes, $\kappa=\kappa_{\rm in}+\kappa_{\rm out}$. For all the allotropes, $\kappa_{\rm in}$ dominates, which means that flexural phonons are not the major heat carrier in phosphorene. Among the three allotropes, \gls{blueP} has the highest thermal conductivity $128 \pm 3$ $\mathrm{Wm^{-1}K^{-1}}$, which is about two orders of magnitude higer than that in \gls{violetP} ($2.36 \pm 0.05$ $\mathrm{Wm^{-1}K^{-1}}$). While these two allotropes is isotropic for in-plane heat transport, \gls{blackP} exhibits strong anisotropy, with $\kappa$ in the zigzag direction ($78.4 \pm 0.4$ $\mathrm{Wm^{-1}K^{-1}}$) being about six times as large as that in the armchair direction ($12.5 \pm 0.2$ $\mathrm{Wm^{-1}K^{-1}}$). Both the strong anisotropy and the magnitudes of the thermal conductivity in \gls{blackP} and \gls{blueP} are well consistent with the existing \gls{bte}-\gls{ald} predictions based on \gls{dft}-based force constants  \cite{Zhu2014PRB,Qin2015PCCP,Liu2015Nanoscale,Zheng2016PRB,Zhang2017SR,Liu2018APL,Zhu2020PhysicaE}. While the simultaneous description of two or more distinct structures is generally beyond the capability of empirical potentials, the results here suggest that our \gls{nep} model has this capability. For the most complex allotrope, \gls{violetP}, the \gls{bte}-\gls{ald} approach is computationally infeasible and we thus have a prediction for its thermal conductivity for the first time. As suggested by the phonon band picture, its magnitude is among the smallest found for elementary crystals.  

\begin{figure}[htb]
\begin{center}
\includegraphics[width=\columnwidth]{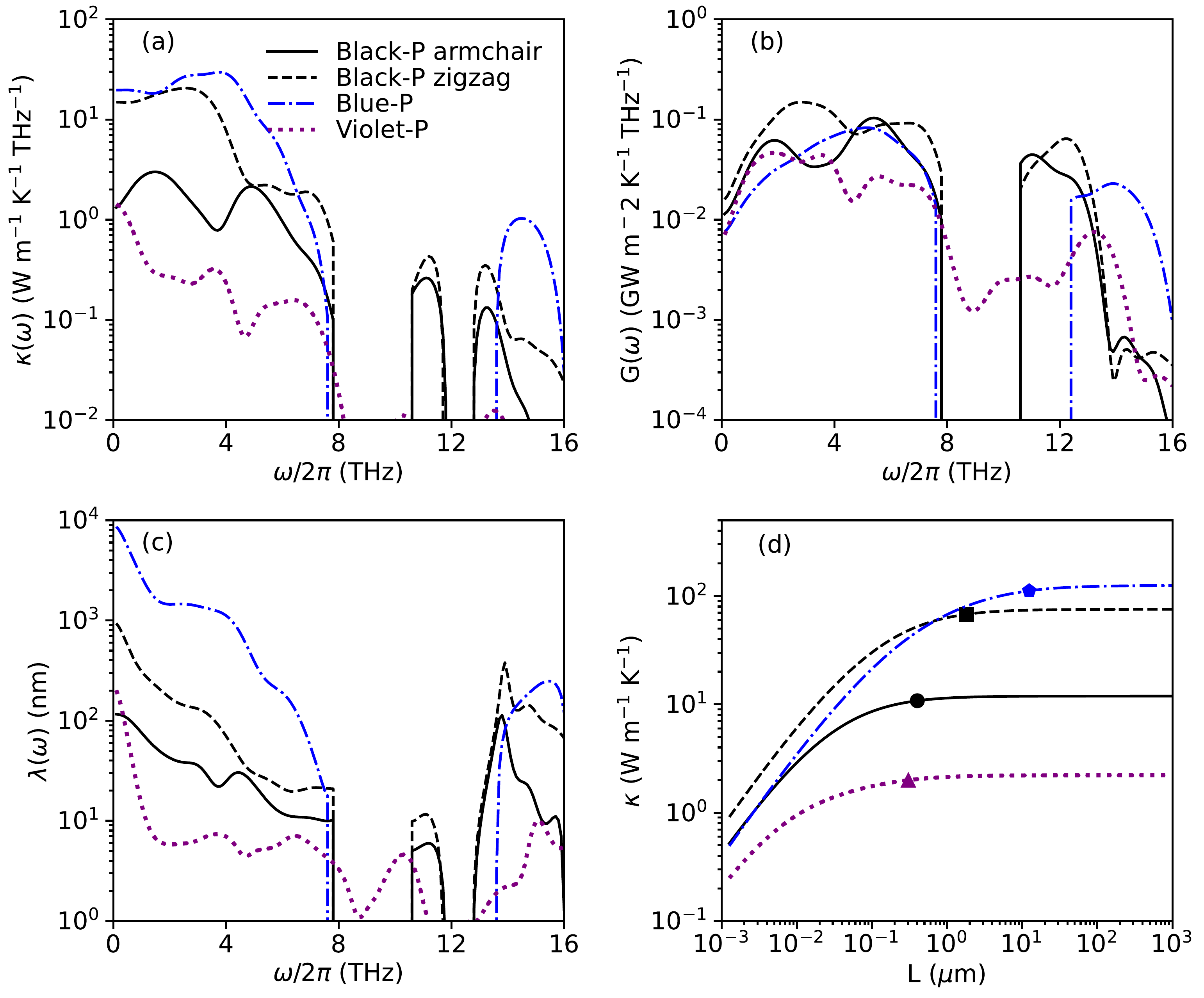}
\caption{ (a) Diffusive spectral thermal conductivity $\kappa(\omega)$, (b) ballistic spectral thermal conductance $G(\omega)$, (c) phonon mean free path $\lambda(\omega)$, and (d) length-dependent thermal conductivity $\kappa(L)$ for the phosphorene allotropes at 300 K and zero pressure. The symbols in (d) denote the system lengths needed to reach 90\% of the convergent $\kappa$ for each allotrope.}
\label{figure:SHC}
\end{center}
\end{figure}

To get more insight in the physical mechanisms behind the highly variable phonon transport in 2D phosphorene, we used the spectral decomposition techniques in Refs.~\onlinecite{Fan2019PRB, Gabourie2021PRB} to calculate the diffusive spectral thermal conductivity $\kappa(\omega)$ and the ballistic spectral thermal conductance $G(\omega)$, as shown in \autoref{figure:SHC}(a) and \autoref{figure:SHC}(b), respectively.
From $\kappa (\omega)$ and $G(\omega)$, one can readily calculate the frequency-dependent phonon \gls{mfp} \cite{Fan2019PRB}:
\begin{equation}
\label{equation:decompose_lambda}
\lambda(\omega)=\kappa(\omega)/G(\omega),
\end{equation}
as shown in \autoref{figure:SHC}(c).
Then, one can obtain the thermal conductivity at any length $L$ \cite{Fan2019PRB},
\begin{equation}
\label{equation:k_L}
\kappa(L)=\int_{0}^{\infty}\frac{d\omega}{2\pi} \frac{\kappa(\omega)L}{L+\lambda(\omega)},
\end{equation}
as shown in \autoref{figure:SHC}(d).

The phonon band gaps in \autoref{figure:phonon} are also manifested in the spectral quantities here. For all the allotropes, $\kappa$ gets it main contribution from the acoustic phonon branches. The maximum phonon \gls{mfp} $\lambda_{\rm max}$ in all the allotropes is located at the low-frequency limit, reaching about $10^4$ nm in \gls{blueP}, $10^3$ nm in the zigzag direction of \gls{blackP}, and  $10^2$ nm in  \gls{violetP} and the armchair direction of \gls{blackP}. These results are well consistent with our choice of the driving force parameter $F_{\rm e}$ that ensures $F_{\rm e}\lambda_{\rm max} \lesssim 1$. Consistent with the different \glspl{mfp}, the allotropes exhibit different convergence rates of $\kappa(L)$ with increasing $L$. It requires more than 10 microns to reach 90\% of the diffusive $\kappa$ in \gls{blueP}, while only requires a few hundred nanometers to reach 90\% of the diffusive $\kappa$ in \gls{violetP}. 

\begin{figure}[htb]
\begin{center}
\includegraphics[width=1.0\columnwidth]{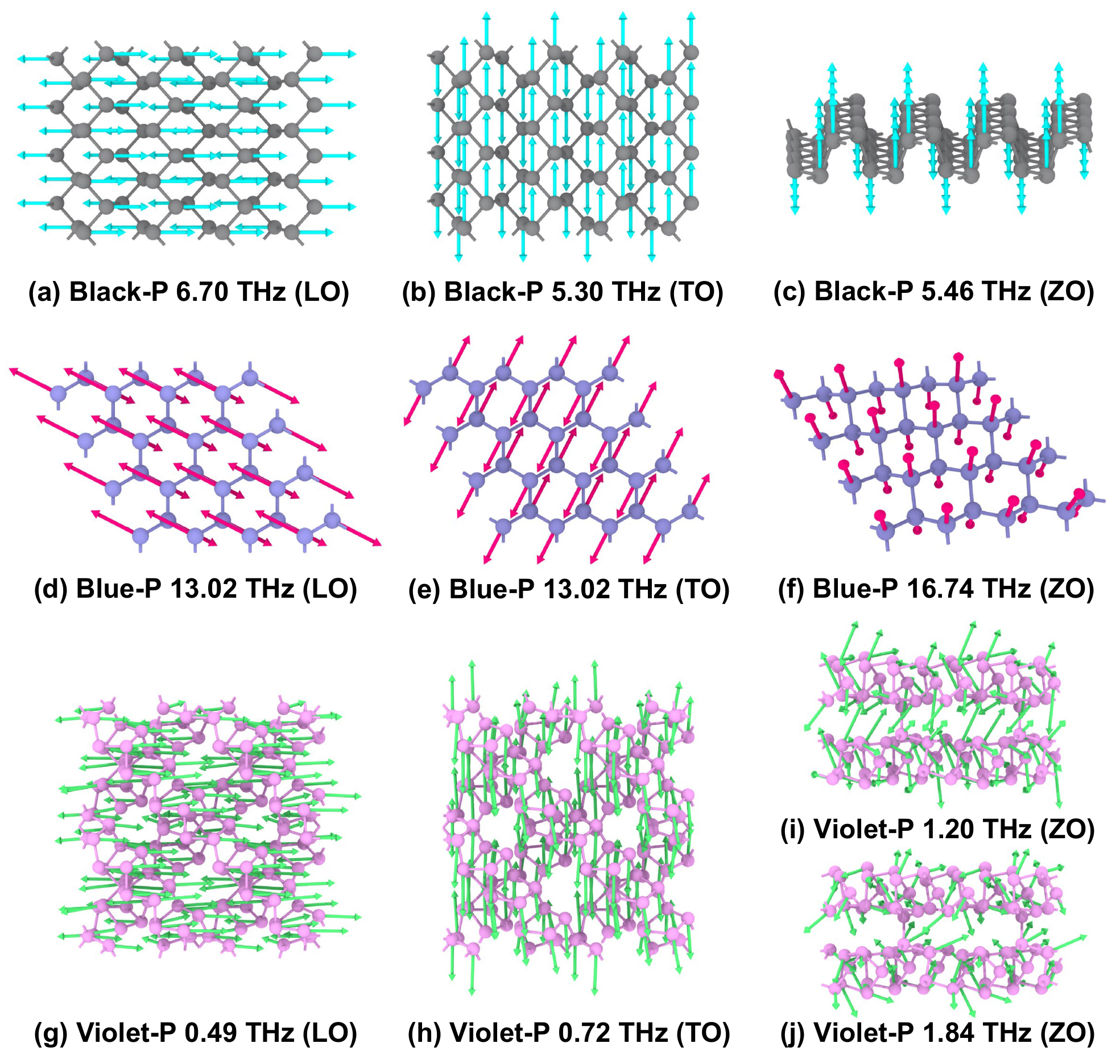}
\caption{Phonon eigenmodes for (a)-(c) \gls{blackP}, (d)-(f) \gls{blueP}, and (g)-(j) \gls{violetP} at selected frequencies $\omega/2\pi$ at the $\Gamma$ point. The magnitude and direction of an arrow represent the eigenvector component at an atom. LO, TO, and ZO represent the longitudinal, transverse, and out-of-plane optical phonon modes, respectively. The \textsc{ovito} package \cite{stukowski2009visualization} is used for visualization.}
\label{figure:eigenmode}
\end{center}
\end{figure}

From \autoref{figure:SHC}(c), we see that the phonon \glspl{mfp} in \gls{violetP} are smaller than 10 nm for $\omega/2\pi \gtrsim 1$ THz. This indicates that most of the phonon modes in \gls{violetP} are localized, which can be confirmed by the optical phonon eigenmodes at the $\Gamma$ point shown in \autoref{figure:eigenmode}, obtained using the method in Ref. \onlinecite{liang2022phononhydrodynamic}. In both \gls{blackP} and \gls{blueP}, the optical eigenmodes at relatively high frequencies exhibit collective movements of the atoms. In contrast, the optical eiginmodes in \gls{violetP} only show collective behavior below 1 THz, and random movements of the atoms starting from 1.2 THz. This indicates that the phonon modes in \gls{violetP} are well localized for $\omega/2\pi \gtrsim 1$ THz.

\begin{figure}[htb]
\begin{center}
\includegraphics[width=1.0\columnwidth]{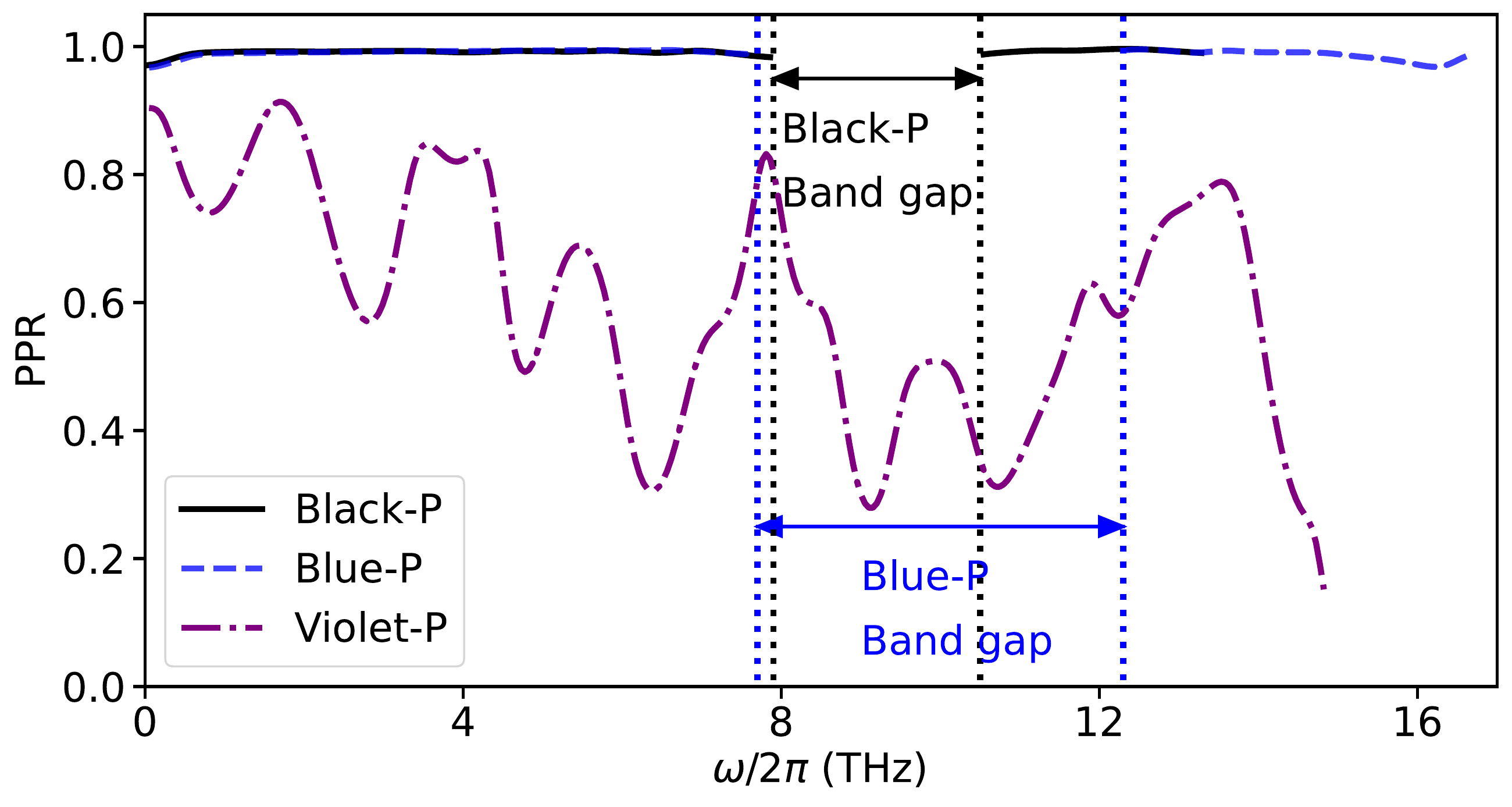}
\caption{Phonon participation ratio as a function of phonon frequency for \gls{blackP}, \gls{blueP}, and \gls{violetP} at 300 K and zero pressure. }
\label{figure:PPR}
\end{center}
\end{figure}

\begin{figure}[htb]
\begin{center}
\includegraphics[width=\columnwidth]{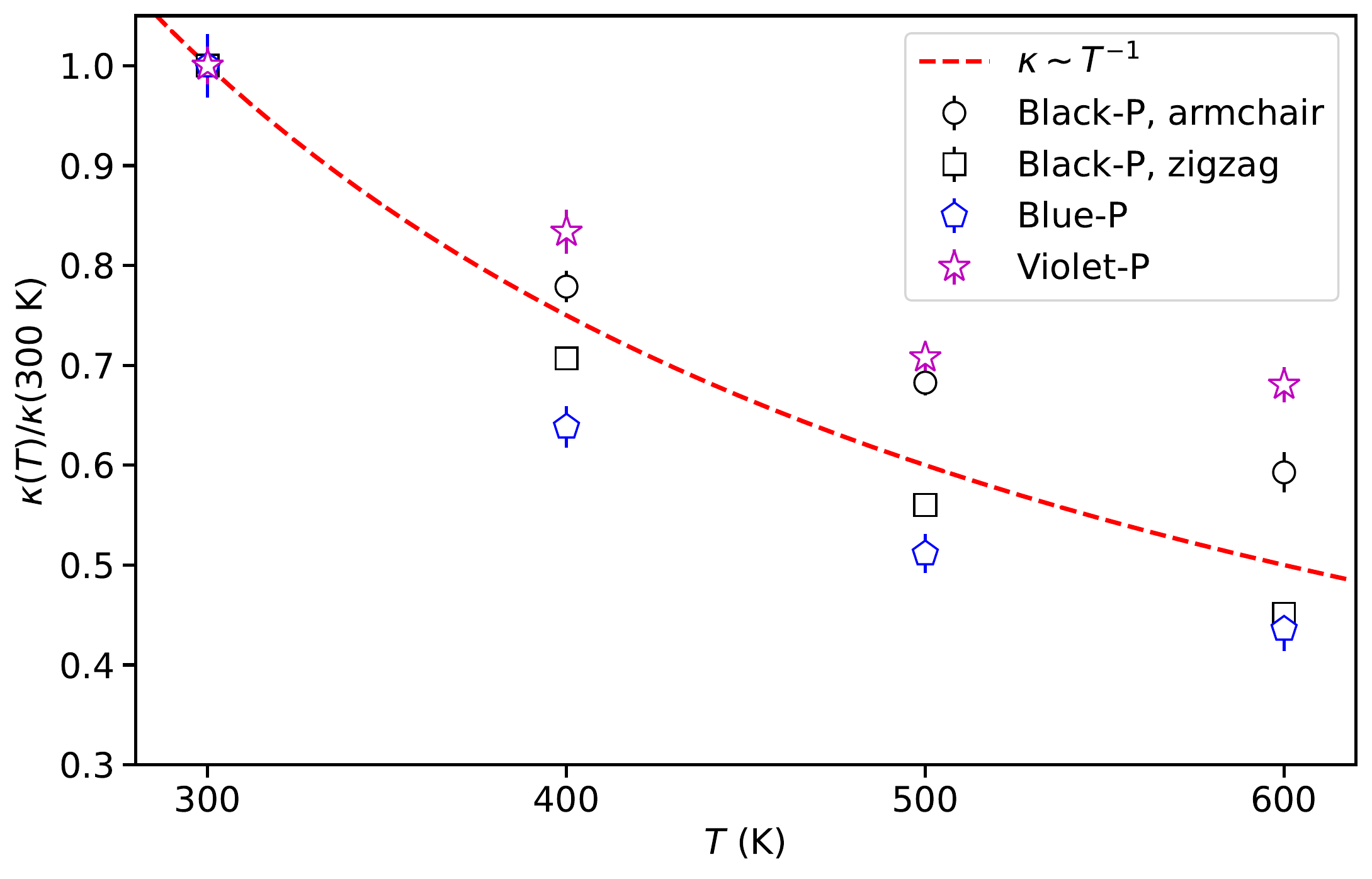}
\caption{Thermal conductivity $\kappa$ for the phosphorene allotropes as a function of temperature $T$. For each allotrope, $\kappa$ is normalized by its value at 300 K.}
\label{figure:temperature}
\end{center}
\end{figure}

Phonon localization can be quantified by the phonon participation ratios (PPR) \cite{burkov1996phonon} defined as
\begin{equation}
\label{equation:ppr}
R(\omega) = \frac{1}{N} 
\frac{\left( \sum_{i} \rho_{i}^{2}(\omega) \right)^2}
{\sum_{i}\rho_{i}^{4}(\omega)},
\end{equation}
where $\rho_i(\omega)$ is the phonon density of states of atom $i$ and $N$ is the total number of atoms involved in the calculation. A value of $R=1$ represents a totally de-localized mode, and a smaller value corresponds to a stronger localization. \autoref{figure:PPR} shows that the phonon modes in \gls{blackP} and \gls{blueP} are essentially de-localized up to the maximum frequency, while those in \gls{violetP} are strongly localizated and its PPR generally decreases with increasing phonon frequency. The strong localization of phonons in \gls{violetP} is consistent with the relatively flat phonon bands and the low phonon group velocities (\autoref{figure:phonon}(c)). 

In \autoref{figure:temperature} we show the temperature-dependent $\kappa$ for the three phosphorene allotropes. \gls{blackP} and \gls{blueP} largely follow a typical $T^{-1}$ dependence of $\kappa$ as dominated by three-phonon scattering processes. However, \gls{violetP} exhibits a clearly weaker temperature dependence, $\kappa \sim T^{-0.59}$, which suggests the importance of high-order anharmonicity as in low-$\kappa$ materials \cite{zeng2021ultralow,zeng2022critical}

\begin{figure}[htb]
\begin{center}
\includegraphics[width=\columnwidth]{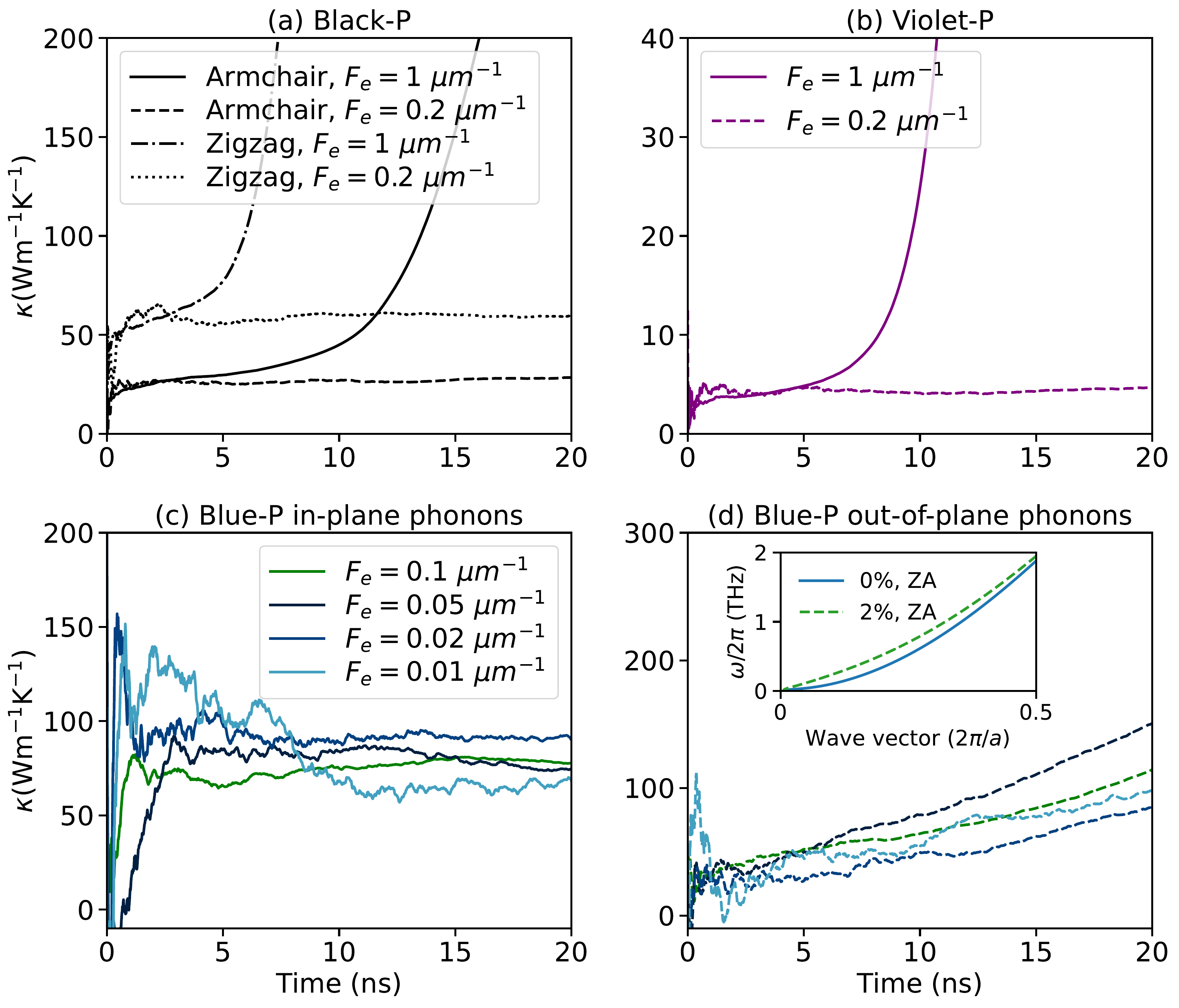}
\caption{
  The thermal conductivity as defined in \autoref{equation:kappa_cumulative} for the phosphorene allotropes at 300 K and different in-plane strain levels: (a) \gls{blackP} under 4\% uniaxial tensile strain, (b) \gls{violetP}, (c)-(d) \gls{blueP} under 2\% biaxial tensile strain. The inset in (d) shows the change of the ZA branch of \gls{blueP} along the $\Gamma$-M path upon the application of 2\% bi-axial tensile strain.
 }
\label{figure:strain_effect}
\end{center}
\end{figure}

An interesting observation in graphene \cite{Pereira2013PRB,Fan2017PRb} and graphene-like \gls{2D} materials \cite{kuang2016nanoscale,xie2016large,li2017thermal,raeisi2019modulated,D2CP01513G} is that the thermal conductivity might be not upper bounded or increase significantly under external strain. Here we use the \gls{hnemd} method to examine this issue for the phosphorene allotropes. For \gls{blackP}, we apply a 4\% uniaxial tensile strain in the armchair or zigzag direction. For \gls{blueP} and \gls{violetP}, we apply 2\% biaxial tensile strain. Our augment below relies on the criteria of $F_{\rm e}\lambda_{\rm max} \lesssim 1$ for keeping the system within the linear-response regime in the HNEMD simulations. In this case, the $\kappa(t)$ defined in \autoref{equation:kappa_cumulative} should converge in the long-time limit. In all the cases we perform \gls{hnemd} simulations up to a long time of 20 ns, which is needed to clearly identify possible $\kappa(t)$ divergence.

The $\kappa(t)$ as defined in \autoref{equation:kappa_cumulative} in the strained systems are shown in \autoref{figure:strain_effect}. For \gls{blackP} and \gls{violetP}, using the value $F_{\rm e}=1$ $\mu$m$^{-1}$ as adopted in the unstrained condition leads to divergent $\kappa(t)$, but the convergence of $\kappa(t)$ can be restored by reducing $F_{\rm e}$ to 0.2 $\mu$m$^{-1}$. This means that $\lambda_{\rm max}$ increases in \gls{blackP} and \gls{violetP} under the external strain but is still finite and $\kappa$ is thus still finite. 

The situation is notably different in \gls{blueP}, which exhibits divergent $\kappa(t)$ even if $F_{\rm e}$ is reduced from 0.1 to 0.01 $\mu$m$^{-1}$. This means that $\lambda_{\rm max}$ in 2\% bi-axially stretched \gls{blueP} is at least $100$ $\mu$m, which in turn means that $\kappa(L)$ does not converge before the millimeter length scale. We cannot indefinitely reduce $F_{\rm e}$ to probe a possible upper bound of $\lambda_{\rm max}$ because of the reduced signal-to-noise ratio with decreased $F_{\rm e}$ in the HNEMD simulations. Nevertheless, our results here do not show a sign of convergence trend of $\kappa(t)$ and we conclude that the thermal conductivity in 2\% bi-axially stretched \gls{blueP} appears unbounded.

The out-of-plane phonons are responsible for the increased $\lambda_{\rm max}$ under stretching (\autoref{figure:strain_effect}(d)). This is in turn related to the linearization of the ZA phonon dispersion around the $\Gamma$ point as shown in the inset of \autoref{figure:strain_effect}(d). Indeed, in the case of silicene, a buckled \gls{2D} material similar to \gls{blueP}, a linear dispersion of the ZA branch can lead to divergent $\kappa$ even in the unstrained condition, according to the \gls{bte}-\gls{ald} approach \cite{gu2015jap,gu2018rmp}. Our findings for strained \gls{blueP} here are consistent with previous \gls{bte}-\gls{ald} results \cite{kuang2016nanoscale,xie2016large,li2017thermal,raeisi2019modulated,D2CP01513G} on strained graphene-like materials, although only three-phonon scattering processes were considered in the \gls{bte}-\gls{ald} approach. Our MD simulations demonstrate the diverse effects of external strain on the thermal conductivity of general \gls{2D} materials in the non-perturbative regime.

\section{Summary and conclusions \label{section:summary}}

In summary, we have constructed a unified \gls{mlp} for three \gls{2D} phosphorene allotropes, \gls{blackP}, \gls{blueP}, and \gls{violetP}, based on the \gls{nep} model \cite{Fan2021PRB,fan2022improving,fan2022GPUMD} that has a comparable accuracy to the existing \gls{gap} model \cite{Deringer2020NC} and a far superior computational efficiency. With this \gls{nep} model, we performed large-scale \gls{md} simulations to study thermal transport in these phosphorene allotropes. For \gls{blackP} and \gls{blueP}, our predicted thermal conductivity based on \gls{hnemd} simulations are consistent with literature results based on the \gls{bte}-\gls{ald} approach, and for \gls{violetP}, our approach allowed for the prediction of its thermal conductivity for the first time. We found that \gls{violetP} has a much smaller thermal conductivity than \gls{blackP} and \gls{blueP}, due to the strong phonon localization in this material. Finally, we find that, under external tensile strain, the thermal conductivity in \gls{blackP} and \gls{violetP} are still finite, but that in \gls{blueP} is still not convergent at least up to the millimeter length scale, due to the linearization of the flexural phonon dispersion. 

\vspace{0.5cm}
\noindent{\textbf{Data availability:}}

Complete input and output files for the \gls{nep} training and testing are freely available at a Zenodo repository \cite{Penghua2022Zendo}. 

\begin{acknowledgments}
We thank Xin Wu, Yanzhou Wang, and Zezhu Zeng for insightful discussions. P.Y. and Z.Z. acknowledge the supports from the National Key R\&D Program of China (No. 2018YFB1502602) and the National Natural Science Foundation of China (Nos. 11932005 and 11772106). T.L. and J.X. acknowledge the support from the Research Grants Council of Hong Kong (Grant No. AoE/P-02/12).
Z.F. acknowledges support from the National Natural Science Foundation of China (No. 11974059). 
T.A-N. has been supported in part by the Academy of Finland through its QTF Centre of Excellence program (No. 312298) and Technology Industries of Finland Centennial Foundation Future Makers grant.
\end{acknowledgments}
\bibliography{main.bib}
\end{document}